\documentclass[aps,amsmath,amssymb,twocolumn]{revtex4-1}

\usepackage{graphicx}
\usepackage{hyphenat}
\usepackage{siunitx}
\usepackage{textgreek}

\usepackage{flushend}

\usepackage[caption=false]{subfig} 

\usepackage{color}

\begin{document}

\title{The thick-crystal regime in photon pair sources}

\author{B. Septriani}
\author{J. A. Grieve}
\author{K. Durak}
\author{A. Ling}
\email[Corresponding author: ]{cqtalej@nus.edu.sg}

\affiliation{Centre for Quantum Technologies, National University of Singapore\\Block S15, 3 Science Drive 2, 117543 Singapore}

\date{\today}


\begin{abstract}
We present comprehensive measurement data on the pump and collection beam parameters necessary to achieve high collection efficiency ($89.0\,\pm\,$\SI{1.7}{\%}) together with high brightness when a single \textbeta-Barium Borate crystal is operated in the thick-crystal regime and pumped with a narrow linewidth laser source. Spectral analysis of the collinear, non-degenerate photons suggest that the effective interaction length within the crystal is dominated by the collection beam mode and the use of longer crystals with increased spatial walk-off does not necessarily lead to a reduced collection efficiency. This result is an important consideration for optical designers who seek to develop practical photon pair sources.
\end{abstract}

\maketitle

Photon pair sources based on spontaneous parametric downconversion (SPDC) have been widely used for the past two decades of quantum optics research \cite{hong86} and continue to play a prominent role in modern experiments \cite{christensen13, giustina13, giovannini15}. Early experiments often employed critically phase matched bulk crystals such as $\beta$-Barium Borate (BBO) \cite{kwiat95,kwiat99}. More recently there has been a trend toward sources employing engineered quasi-phase matched materials \cite{qpmshg} such as PPKTP \cite{fiorentino05}. 

SPDC source designs, although highly successful for in-lab demonstrations of fundamental physics or proof-of-principle technology demonstrations, are often unsuitable for field deployment in applications where the requirements of size, weight and power (SWAP) are very strict~\cite{tang14, chandrasekara15_spie2}. In addition, it is necessary to consider the overall brightness and collection efficiency (observed pair-to-singles ratio) into single-mode optics~\cite{kurtsiefer01,ling08}, for applications such as device-independent quantum communication~\citep{acin07}.

To meet the SWAP requirements, it is advantageous to have a bright and efficient SPDC source that generates collinear emission (minimizing size) while being able to operate without active temperature control (minimizing weight and power and improving ruggedness). A design based on Type I, non-degenerate SPDC that meets the above requirements has previously been reported \cite{trojek08} to exhibit single-mode brightness and efficiency competitive with sources based on quasi-phase matching \cite{dixon14}. 

Relying on long crystals (\SI{15.76}{\mm}), the reported design was clearly operating in the thick-crystal regime of SPDC \cite{sutherlandbook}, where the interaction length is defined only by the overlap of the optical beams. This is a regime that is not commonly used for photon pair sources because the impact of spatial walk-off on the collection efficiency and brightness is not well understood. Most reports deal with the thin-crystal regime where the crystal facets define the interaction length, by using non-critical phase matching (such as with PPKTP~\cite{dixon14}) or with very thin angle-tuned crystals~\citep{christensen13}. However, the fact that thick angle-tuned crystals are simpler to fabricate and handle (with knock-on effects for instrument cost) makes it interesting to understand the optimal operating parameters for the thick-crystal regime in order to make meaningful trade-offs when designing a practical photon pair source.

\begin{figure}[!htb]
    \centering
    \includegraphics[width=0.9\linewidth]{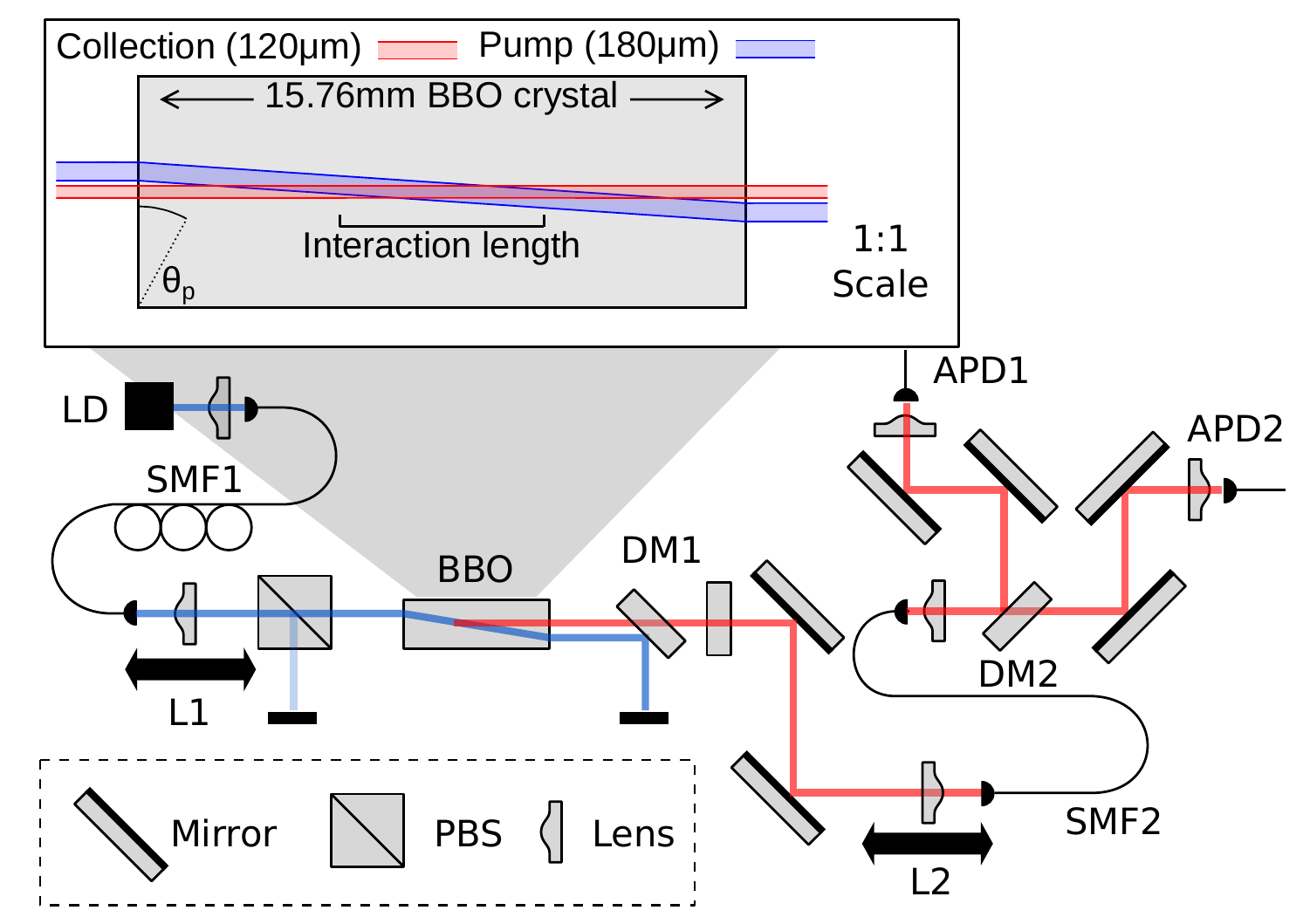}
    \caption{\label{fig:scheme} The experiment setup for testing the thick-crystal regime (see inset). Pump and collection beam focal sizes are adjusted by translation of lenses L1 and L2, with optical paths adjusted so that the focii overlap at the center of the BBO crystal. The pump beam spatial mode is filtered by a single mode fiber (SMF1), and SPDC emission is collected into another single-mode fiber (SMF2). Pump, signal and idler beams are separated by dichroic mirrors (DM1, DM2). Photon pairs are detected by fiber-coupled avalanche photodiodes (APD1, APD2). Measured transmission of the setup is reported in Table.~\ref{table:eff}.}
\end{figure}

In this paper, we study the SPDC emission in the thick-crystal regime where the crossing of the optical beams results in an effective interaction length that lies within the crystal, as illustrated in Fig.~\ref{fig:scheme}. We show that the reduced interaction length leads to a broadened SPDC spectrum. In addition, we have performed a systematic study to map the range of useful pump and collection beam conditions for optimizing brightness and collection efficiency.

The experiment setup is shown in Fig.~\ref{fig:scheme}. The pump source is a \SI{405}{\nm} laser diode with an integrated volume holographic grating that enables a laser linewidth of approximately \SI{150}{\MHz}. The diode output is coupled into a single-mode fiber to  clean the spatial profile of the pump beam before being launched towards a BBO crystal. Throughout the experiment, the focii of the pump and the collection beams overlap at the crystal's mid-point. The beam size and the focus position are determined with a CCD camera, and checked after each realignment to ensure the beam overlap is consistent throughout the whole data set.

The generated photon pairs are collected into a single-mode fiber before being separated by a dichroic beam splitter. The photons are registered by passively quenched Si avalanche photodiodes (APDs). The BBO crystal (of length \SI{15.76}{\mm}) was tuned to generate collinear, non-degenerate photon pairs centered on \SI{780}{\nm} (signal) and \SI{842}{\nm} (idler). The system detection probabilities for the signal and idler wavelengths were determined to be $37.4\,\pm$\,\SI{1.5}{\%} and $33.0\,\pm\,$\SI{1.4}{\%} (see Table~\ref{table:eff}), taking into account the contributions from measured losses at each optical element and the wavelength dependent detector efficiency.

\begin{figure*}[t!]
    \centering
    \includegraphics[width=0.8\linewidth]{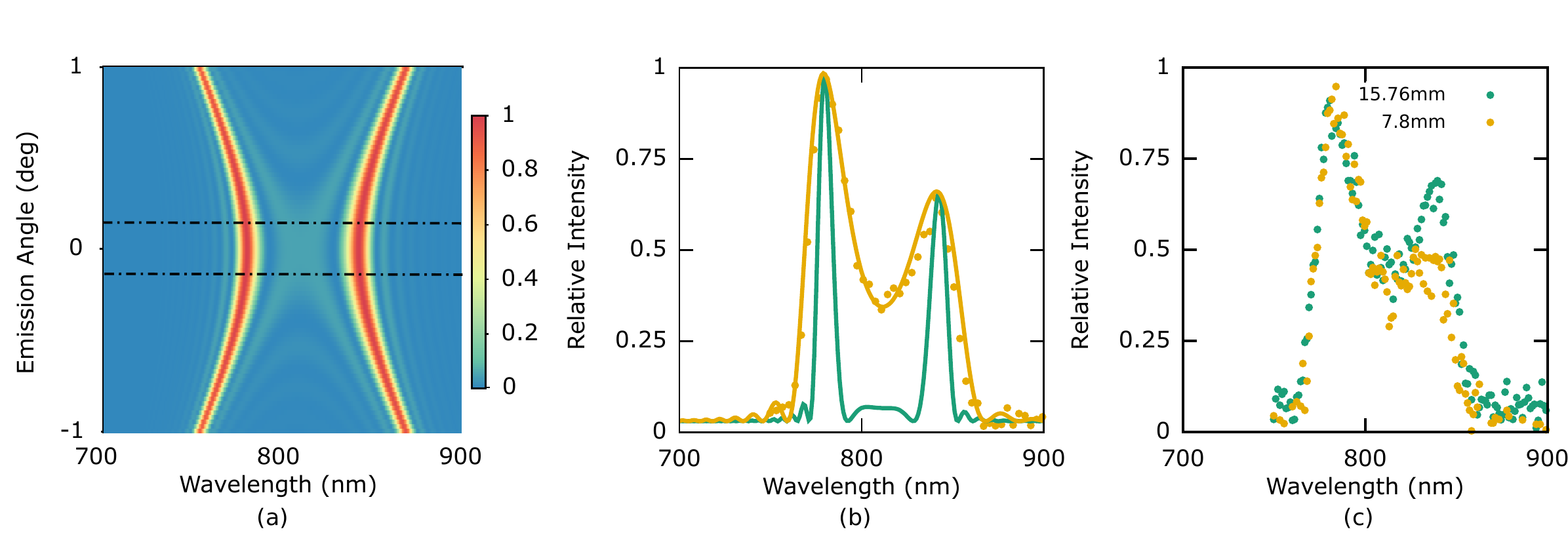}
    \caption{\label{fig:js} Predicted and observed spectra for collected photon pairs. The pump focal FWHM is \SI{180}{\micro m}. (a) The emission angle for different wavelengths for collinear, non-degenerate SPDC using a \SI{15.76}{\mm} length BBO crystal. The dashed lines represent the acceptance angles for a collection beam with a focal FWHM of \SI{120}{\micro m} targeting \SI{780}{\nm}. (b) The observed spectrum (points) fits the predicted emission from a \SI{5.35}{\mm} interaction length and is much broader than the prediction assuming the crystal length (green line). (c) The observed spectra for two different crystal lengths (\SI{15.76}{\mm} and \SI{7.8}{\mm}) under identical pump and collection beam parameters. The spectral width are identical despite a reduction in crystal length. This supports the hypothesis that the effective interaction length in the thick-crystal regime is dominated by the crossing of the optical beams.}
\end{figure*}

The effective interaction length of the setup was investigated using the spectrum of the collected SPDC photons. The SPDC emission angles for different wavelengths are shown in Fig.~\ref{fig:js}(a). This heat map illustrates the peak emission angle for a given wavelength when the BBO crystal is tuned to emit \SI{780}{\nm} light in the collinear direction. The heat map assumes that the pump has a narrow linewidth and is a plane wave. Both conditions are approximated well by using the grating-stabilized laser diode and a pump whose focal FWHM is $180.0\,\pm\,$\SI{3.8}{\micro m}, corresponding to a Rayleigh range of approximately $181.6\,\pm\,$\SI{3.8}{\mm} with both values measured in air.

The expected spectrum after coupling into a single-mode fiber can be obtained by a numerical integration of the heat map across the acceptance angle of the collection beam \cite{kurtsiefer01}. In our experiment the FWHM at the focus is approximately $120.0\,\pm\,$\SI{1.2}{\micro m} with a corresponding (full) acceptance angle of $0.277\,\pm\,$\SI{0.002}{\degree}. This yields the green line in Fig.~\ref{fig:js}(b). The measured spectrum (using a grating spectrometer with \SI{1}{\nm} resolution) is much broader, corresponding to the emission from a \SI{5.35}{\mm} interaction length. 

To explain this discrepancy, we note that the heat map in Fig.~\ref{fig:js}(a) neglects the effect of spatial walk-off. The lateral deviation of the pump beam over a length of \SI{5.35}{\mm} is approximately \SI{368}{\micro m}. This value matches well with the full aperture of the collection beam and supports the hypothesis that in the thick-crystal regime the effective interaction distance is dominated by the crossing of the optical beams, rather than the crystal length. The broadened spectrum that we report here is technically different from a previous observation in~\cite{trojek08}, which was caused by the use of a free-running pump laser~\cite{trojekphd} masking the effect of the thick-crystal regime. 

\begin{figure*}[t] 
    \centering
    \includegraphics[width=0.8\linewidth]{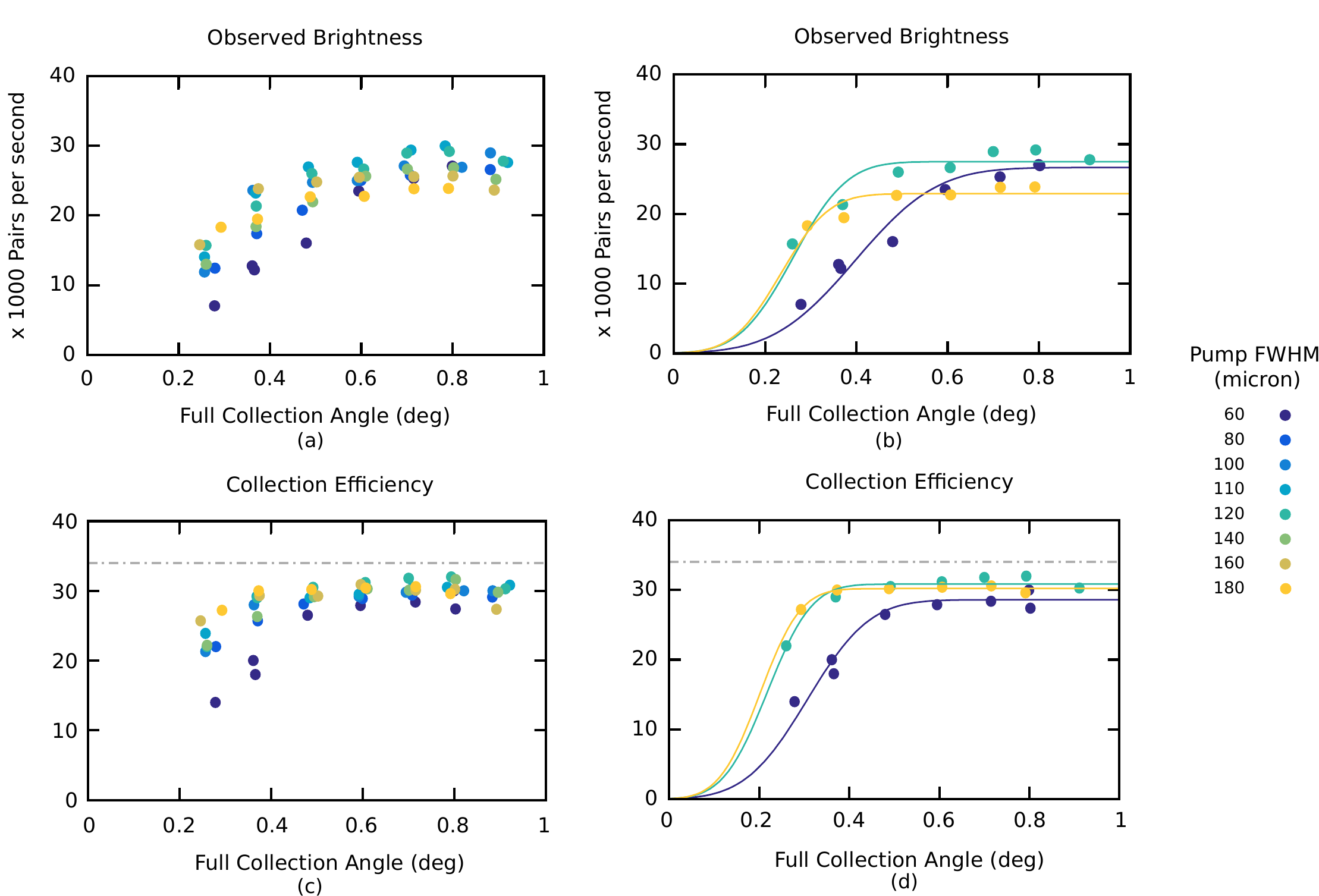}
    \caption{\label{fig:combo} The brightness and collection efficiency observed as pump and collection beam parameters are tuned (see key for range of pump sizes). (a) Observed brightness per mW of pump power for the full set of collected data. (b) Data from (a) for three selected pump sizes (\SI{60}{\micro m}, \SI{120}{\micro m} and \SI{180}{\micro m}) with their respective fit to the error function (solid lines). (c) Collection efficiency for the full set of collected data. (d) Data from three selected pump sizes (\SI{60}{\micro m}, \SI{120}{\micro m} and \SI{180}{\micro m}) repeated from (a) with their respective fit to the error function (solid lines). The dashed lines represent the measured system efficiency, with the observed collection efficiency typically approaching 90\% of this value. We note that error bars are too small to be significant. It is also interesting to note the reduced spread in the efficiency data when considered alongside the corresponding brightness values. This implies that uncertainty in the latter is dominated by power fluctuations (typically \SI{0.8}{\%}), while the efficiency (as a ratio) is affected only by the quality of alignment.}
\end{figure*}

The optical beam parameters were then fixed, and a shorter (\SI{7.88}{\mm}) crystal was placed at the beam overlap. The resulting spectrum is presented in Fig.~\ref{fig:js}(c) and we note that a reduction in crystal length by half did not change the observed emission bandwidth. This is consistent with the hypothesis presented above.

A practical photon pair source must exhibit high brightness without sacrificing efficiency. We performed a systematic investigation of the performance achievable with this geometry. With the original \SI{15.76}{\mm} length crystal restored, the pump focal FWHM was systematically increased from \SI{60}{\micro m} to \SI{180}{\micro m}. For each pump size the brightness and collection efficiency were recorded for a range of collection angles. We identify the collection efficiency of the signal photons, $\eta_s$, as the ratio of pairs to singles, $\eta_s = C/S_i$, where $C$ is the observed pair rate and $S_i$ is the rate of idler photons. The brightness and efficiency data are presented in Fig.~\ref{fig:combo}. 

The highest values for efficiency and brightness occurred for pump modes of \SI{110}{\micro m} and \SI{120}{\micro m}. The maximum observed value for $\eta_s$ is $33.6\,\pm\,$\SI{0.4}{\%}  (after correcting for the dark count rate of the detectors), $\approx$ 89\% of the maximum system efficiency (see Table~\ref{table:efffinal}). This implies a coupling efficiency into the fiber mode  of at least $89.0\pm\,$\SI{1.7}{\%} for the \SI{780}{\nm} photons. The observed photon pair rate at this value was $29,200\,\pm\,224$ pairs per second per mW of pump power, with the error dominated by laser power fluctuations of \SI{0.8}{\%}. These values are consistent with previously reported pair sources \cite{trojek08}, with additional data regarding scaling of brightness and efficiency against beam parameters.

\begin{table}[b]
    \begin{tabular}{p{2.5cm}| p{2.5cm} | p{2.5cm}}
    Element & Signal (\SI{780}{\nm}) & Idler (\SI{842}{\nm}) \\
    \hline
    BBO crystal & $98.77\,\pm\,$\SI{0.01}{\%} & $98.90\,\pm\,$\SI{0.04}{\%} \\
    Optical Elements & $76\,\pm\,$\SI{3}{\%} & $72\,\pm\,$\SI{3}{\%} \\
    3 Fiber Interfaces & \SI{88}{\%}  (estimated) & \SI{88}{\%} (estimated)\\
    Detector & $57.2\,\pm\,$\SI{0.8}{\%} & $52.5\,\pm\,$\SI{0.6}{\%}\\ \hline
    System Efficiency & $38\,\pm\,$\SI{1}{\%} & $33\,\pm\,$\SI{2}{\%}
    \end{tabular}
    \caption{\label{table:eff} Transmission and detection efficiency for signal and idler photons at different segments of the experiment setup (see Fig.~\ref{fig:scheme}).}
\end{table}

\begin{table}[b]
    \begin{tabular}{p{1.5cm}|p{1.8cm} |p{1.8cm} | p{1.8cm}}
     Wavelength \newline (nm) & System \newline Efficiency & Observed \newline Efficiency \newline & Actual\newline Efficiency \\
    \hline
    780 & $37.8\,\pm\,$\SI{0.6}{\%} & $33.6\,\pm\,$\SI{0.4}{\%} & $89.0\,\pm\,$\SI{1.7}{\%}\\
    842 & $33.0\,\pm\,$\SI{1.5}{\%} & $27.0\,\pm\,$\SI{0.3}{\%} & $81.9\,\pm\,$\SI{3.7}{\%}
    \end{tabular}
    \caption{\label{table:efffinal} Detector output for the data point with maximum observed efficiency, $\eta_s$ (Fig.~\ref{fig:combo} (d)). The rate of single events for signal and idler photons are $112,400\,\pm\,864$ and $91,150\,\pm\,700$ respectively, while the rate of detected photon pairs is $29,200\,\pm\,224$. The dark count (DC) rate for each detector is approximately 4,400, leading to signal and idler collection efficiencies of $89.0\,\pm\,$\SI{1.7}{\%} and $81.9\,\pm\,$\SI{3.7}{\%} respectively.}
\end{table}

It is remarkable that this efficiency is achieved even when the signal and idler spectra are not clearly distinguished (see Fig.~\ref{fig:js}). The slope of the dichroic mirror used in this experiment transitions between minimum and maximum transmission over \SI{12}{\nm} and is centered on \SI{801}{\nm} and impacts mostly signal photons (wavelength below \SI{810}{\nm}). The result is that about 9\% of signal photons are routed into the idler arm, lowering the rate of observed coincidences and the estimated collection efficiency for signal photons. Meanwhile, the collection of idler photons is not affected (see Supplementary Material for a detailed treatment). Taking this into account, the overlap of the signal photon emission with the collection mode of the fiber is estimated to be nearly 100\%. It should be possible to alter the source to move SPDC emission outside the filter cross-over region, enabling direct observation of this high degree of mode overlap. However, this is left to future work due to the difficulty in obtaining the appropriate alignment lasers for calibrating the optical path.

When considering the imbalance between signal and idler collection efficiency, we note the difficulty in optimizing the collection for both beams simultaneously (the collection beam is aligned using a \SI{780}{\nm} light source). If signal and idler were collected separately, it should be possible to improve the balance between the two arms and hence the overall brightness. Due to the difficulty in obtaining appropriate alignment lasers, this was left for future work. A final source of loss attributed to saturation behaviour of the passively quenched detectors~\cite{grieve15} is minimized by operating in the low-flux regime and then re-normalizing the rates to unit power and time.

Both brightness and efficiency reach a plateau for every pump mode under test. The maximal achievable values are clustered relatively close together for all the different pump sizes. We also represent the data from three pump values with their respective (fitted) trends. The overlap of the SPDC emission pattern (with its almost circular symmetry) with an increasing acceptance cone gives rise to a smooth stepwise transition. As a simple function of this form, we choose the error function to provide a guide to the eye (a more rigorous model is in preparation~\cite{shalmcomm}).  With the aid of these trend lines, we observe that the larger pump sizes can tolerate a larger range of collection angles before the roll-off towards lower brightness and efficiency. This enhanced range tolerance is expected for larger pump sizes as the effective spatial walk-off within the collection volume is reduced.

It is also interesting to note that increasing pump size was associated with a small drop in the maximal brightness, while the maximal collection efficiency was unchanged. For the largest pump beam size in the data (\SI{180}{\micro m}), this drop was statistically significant (at approximately 42 standard deviations, see Fig.~\ref{fig:combo}(b)). We can exclude pump wavefront curvature as a cause since the Rayleigh ranges of all pump beams exceed the crystal length by many multiples. Instead, this observation is explained by noting that for a smaller pump focal FWHM, the SPDC emission between the start and the end points of the interaction regime will have better overlap with the collection beam, compared to SPDC emission that is further separated when using a broader pump.

The overall trends in the observed data for brightness and collection efficiency were very gradual. While there is no clearly superior ``sweet-spot'', it is quite straightforward to identify the range of optimal collection beam parameters. This illustrates the robust nature of the thick-crystal regime, with good performance over a range of collection beam sizes. The fact that both source brightness and collection efficiency have a similar range of optimal collection angles further suggests that photon pairs produced in Type-I collinear SPDC have a good overlap with single-mode collection.

From the perspective of an optical designer, this tolerance of brightness and collection efficiency to a wide range of pump and collection beam parameters makes the thick-crystal regime a good candidate geometry for developing practical and effective photon pair sources. The simplified handling of these longer crystals should also have knock-on effects for fabrication cost in the final instrument designs.

Our data clearly shows that a large spatial walk-off does not necessarily lead to poor collection efficiency and brightness. After taking into account the measured efficiency of all the optical elements, we can conclude that the overlap of the signal photons with the fiber collection mode occurs with almost unit efficiency. Further enhancements to our experiment to increase the collection efficiency of idler photons was also discussed.

Furthermore, we have shown that over the range of practical pump and collection beam parameters tested, the error function provides a good guide to the trend in brightness and collection efficiency. This physically motivated heuristic can be an invaluable design aid. Work is ongoing to test whether entangled photon pair sources built around a pair of thick crystals exhibit similar trends in high brightness and collection efficiency.

This work is supported by the National Research Foundation grant NRF-CRP12-2013-02 \emph{``Space based Quantum Key Distribution''} and the MOE grant MOE2012-T3-1-009 \emph{``Random Numbers from Quantum Processes''}. We thank Tang Zhongkan for spectrometer preparation, Siddarth Joshi for assistance with detector calibration and Alessandro Cere for helpful discussion of the manuscript.


\begin{thebibliography}{10}
\newcommand{\enquote}[1]{``#1''}

\bibitem{hong86}
C.~K. Hong and L.~Mandel, Phys. Rev. Lett. \textbf{56}, 58 (1986).

\bibitem{christensen13}
B.~G. Christensen, K.~T. McCusker, J.~B. Altepeter, B.~Calkins, T.~Gerrits,
  A.~E. Lita, A.~Miller, L.~K. Shalm, Y.~Zhang, S.~W. Nam, N.~Brunner, C.~C.~W.
  Lim, N.~Gisin, and P.~G. Kwiat, Phys. Rev. Lett. \textbf{111}, 130406
  (2013).

\bibitem{giustina13}
M.~Giustina, A.~Mech, S.~Ramelow, B.~Wittmann, J.~Kofler, J.~Beyer, A.~Lita,
  B.~Calkins, T.~Gerrits, S.~W. Nam, R.~Ursin, and A.~Zeilinger,
  Nature \textbf{497}, 227 (2013).

\bibitem{giovannini15}
D.~Giovannini, J.~Romero, V.~Poto\v{c}ek, G.~Ferenczi, F.~Speirits, S.~M.
  Barnett, D.~Faccio, and M.~J. Padgett, Science
  \textbf{347}, 857 (2015).

\bibitem{kwiat95}
P.~G. Kwiat, K.~Mattle, H.~Weinfurter, A.~Zeilinger, A.~V. Sergienko, and Y.~H.
  Shih, Phys. Rev. Lett. \textbf{75}, 4337 (1995).

\bibitem{kwiat99}
P.~G. Kwiat, E.~Waks, A.~G. White, I.~Appelbaum, and P.~H. Eberhard,
  Phys. Rev. A
  \textbf{60}, R773 (1999).

\bibitem{qpmshg}
M.~Fejer, G.~Magel, D.~Jundt, and R.~Byer, IEEE J. Quantum Electron.
  \textbf{28}, 2631 (1992).

\bibitem{fiorentino05}
M.~Fiorentino, C.~Kuklewicz, and F.~Wong, Opt. Express \textbf{13}, 127 (2005).

\bibitem{tang14}
Z.~Tang, R.~Chandrasekara, Y.~Y. Sean, C.~Cheng, C.~Wildfeuer, and A.~Ling,
  Sci. Rep.
  \textbf{4}, 6366 (2014).

\bibitem{chandrasekara15_spie2}
R.~Chandrasekara, T.~Zhongkan, T.~Y. Chuan, C.~Cheng, B.~Septriani, K.~Durak,
  J.~A. Grieve, and A.~Ling, Proc. SPIE
   (2015). 96150S.

\bibitem{kurtsiefer01}
C.~Kurtsiefer, M.~Oberparleiter, and H.~Weinfurter, Phys.
  Rev. A \textbf{64}, 023802 (2001).

\bibitem{ling08}
A.~Ling, A.~Lamas-Linares, and C.~Kurtsiefer, Phys. Rev. A \textbf{77}, 043834 (2008).

\bibitem{acin07}
A.~Acin, N.~Brunner, N.~Gisin, S.~Massar, S.~Pironio, and V.~Scarani,
  Phys. Rev. Lett. \textbf{98}, 230501 (2007).

\bibitem{trojek08}
P.~Trojek and H.~Weinfurter, Appl.
  Phys. Lett. \textbf{92}, 211103 (2008).

\bibitem{dixon14}
P.~B. Dixon, D.~Rosenberg, V.~Stelmakh, M.~E. Grein, R.~S. Bennink, E.~A.
  Dauler, A.~J. Kerman, R.~J. Molnar, and F.~N.~C. Wong, Phys. Rev. A \textbf{90}, 043804 (2014).

\bibitem{sutherlandbook}
R.~L. Sutherland, (Marcel Dekker, Inc.,
  2003), 1st ed. {ISBN: 0-8247-4243-5}.

\bibitem{trojekphd}
P.~Trojek, Ph.D. thesis, LMU (2007).

\bibitem{grieve15}
J.~A. Grieve, R.~Chandrasekara, Z.~Tang, C.~Cheng, and A.~Ling,
  {arXiv:1509.03959 [quant-ph]}  (2015).

\bibitem{shalmcomm}
L.~K. Shalm, Personal communication.


\end{thebibliography}


\clearpage


\section*{Supplementary Material}

\subsection{Losses in the dichroic mirror}

\begin{figure}[h] 
    \centering
    \includegraphics[width=0.9\linewidth]{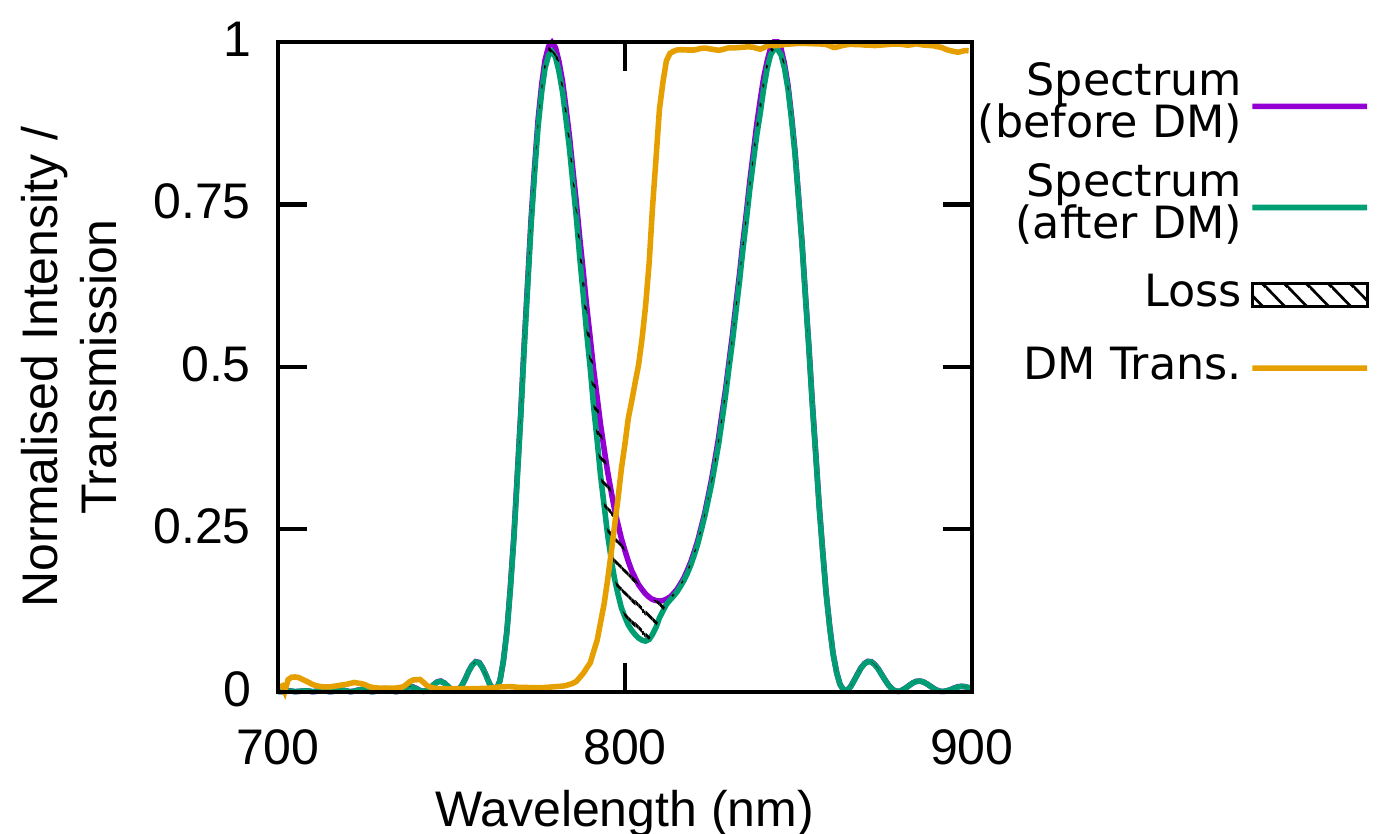}
    \caption{\label{fig:dm-spectrum} Transmission spectrum of the dichroic mirror (FF801-DiO2, Semrock). We also plot the simulated joint spectrum for a \SI{5.35}{\mm} BBO crystal with collection over a \SI{0.3}{\degree} solid angle, with the loss due to misdirected photons indicated.}
\end{figure}

\begin{figure}[h] 
    \centering
    \includegraphics[width=0.35\linewidth]{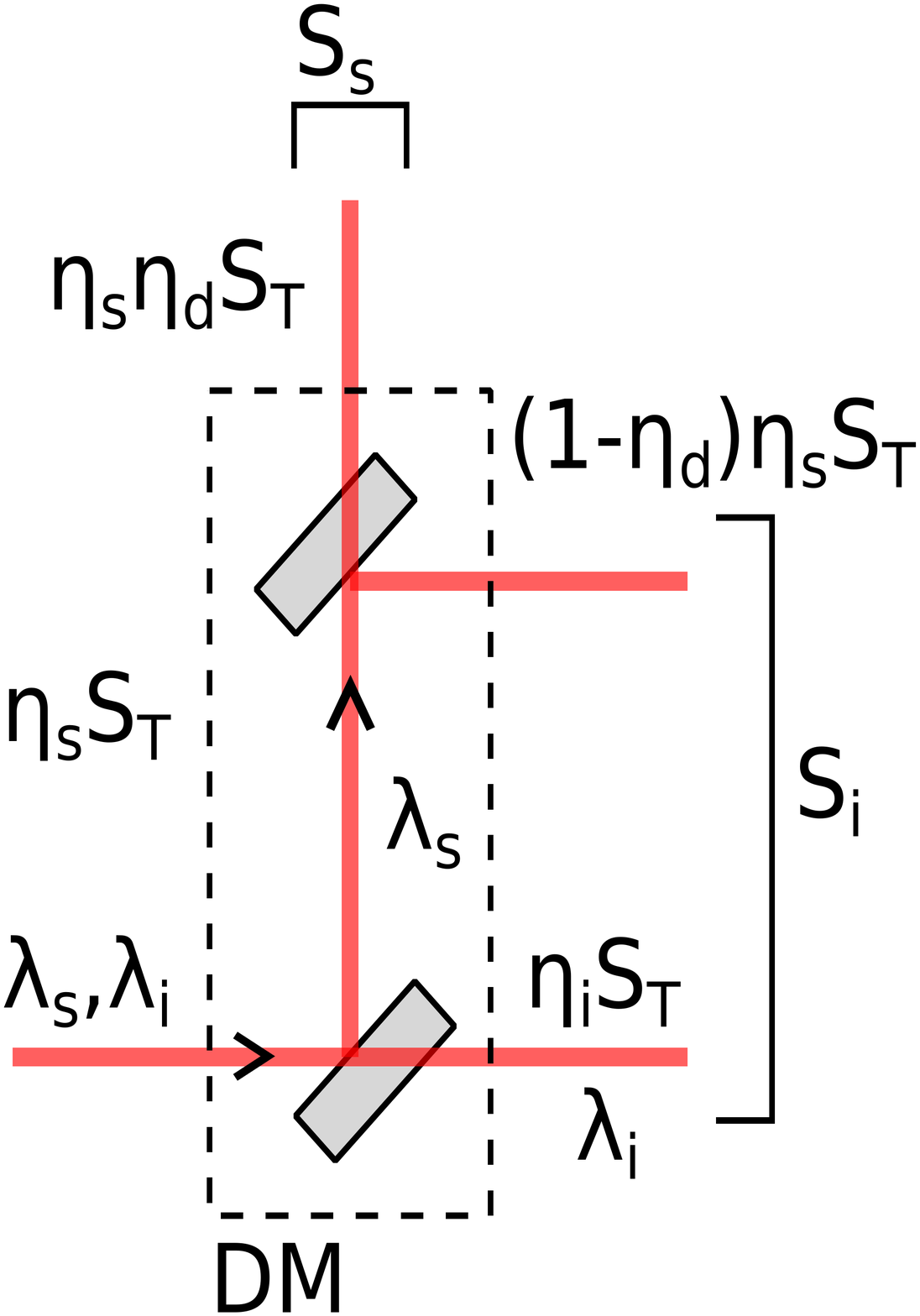}
    \caption{\label{fig:dichroic-logic} A model for the effect of the dichroic mirror on the observed single photon rates.}
\end{figure}

The measured transmission spectrum of the dichroic mirror used to split the signal and idler beams (FF801-DiO2, Semrock, denoted DM2 in Fig.~\ref{fig:scheme}) is plotted in Fig.~\ref{fig:dm-spectrum}. The cut-off is centred at 801\,nm and has a width of roughly 15\,nm. As our signal and idler spectra are centered on 810\,nm and imperfectly separated (Fig.~\ref{fig:js}b), the finite width causes some of the incident photons to be misdirected. As this effect occurs primarily for shorter wavelengths, we treat this approximately by assuming only signal photons are affected. This enables us to model the element as the combination of a perfect dichroic mirror and a partially reflecting mirror placed in the signal arm (Fig.~\ref{fig:dichroic-logic}).

We denote the system transmission efficiency of the signal and idler photons up to the dichroic mirror as $\eta_s$ and $\eta_i$ respectively, and the transmission coefficient of our model's partially reflecting mirror $\eta_d$. The corresponding modified values of the coincidence rate $C$ and the single photon rates $S_i$ and $S_s$ are denoted $C'$, $S'_i$ and $S'_s$. We also consider a ``true rate'' of photons $S_T$, and identify the single photon rates with

\begin{equation}
    \label{eqn:si-ss}
    S_i = \eta_i S_T;\quad
    S_s = \eta_s S_T.
\end{equation}

Considering the observed coincidences in these terms,

\begin{equation}
    \label{eqn:c-prime}
    C' = \eta_s \eta_d \eta_i S_T,
\end{equation}

\noindent we can form an expression for the system transmission efficiency $\eta_i$ and show that it is independent of the value of $\eta_d$:

\begin{equation}
    \label{eqn:eta-i}
    \eta'_i = \frac{C'}{S'_s} = \frac{\eta_d C}{\eta_d S_s} = \frac{C}{S_s}.
\end{equation}

The observed value of $\eta_s$ will be affected. While the coincident rate $C'$ is reduced compared to $C$, the single photon rate $S'_i$ will be boosted by the arrival of misdirected signal photons. These will only trigger an event when the corresponding idler photon is not present, and we ignore the slight increase in detection probability when both are present to write

\begin{equation}
    \label{eqn:Si-prime}
    S'_i = S_i + (1 - \eta_d) (1 - \eta_i) \eta_s S_T.
\end{equation}

Taking this into account, we arrive at the following expression for $\eta_s$

\begin{equation}
    \label{eqn:eta-s}
    \eta_s = \frac{C}{S_i} = \frac{C'}{\eta_d S'_i - (1 - \eta_i)(1 - \eta_d)S'_s}.
\end{equation}

By evaluating the magnitude of the misdirected signal as a fraction of the calculated spectrum (see Fig.~\ref{fig:dm-spectrum}), we can quantify this effect. With $\eta_d \approx 0.91\,\%$ and all other parameters measured, we arrive at a revised system efficiency for $\eta_s$ of \SI{34}{\%}. The highest observed value for $\eta_s$ was $33.6\,\pm\,$\SI{0.4}{\%}, indicating near unit efficiency for the overlap of the signal photons with the fiber collection mode.

\end{document}